\documentclass[preprint,11pt]{elsarticle}

\usepackage{amssymb}
\usepackage{subcaption}

\usepackage{url}

\usepackage[nodots,nocompress]{numcompress}

\journal{Astroparticle Physics}

\begin{document}

\begin{frontmatter}



\title{A Model of the Cosmic Ray Induced Atmospheric Neutron Environment}

\author[label1,label2]{Merlin Kole\corref{cor1}}
\author[label1,label2]{Mark Pearce}
\author[label1,label3]{Maria Mu\~noz Salinas}

\cortext[cor1]{Corresponding author. Tel.: +46 85 537 8186 ; fax: +46 85 537 8216. E-mail address: merlin@particle.kth.se}

\address[label1]{KTH Royal Institute of Technology, Department of Physics, 10691 Stockholm, Sweden}
\address[label2]{The Oskar Klein Centre for Cosmoparticle Physics, AlbaNova University Centre, 10691 Stockholm, Sweden}
\address[label3]{Currently at Ecole Polytechnique, Palaiseau, France}

\begin{abstract}

In order to optimise the design of space instruments making use of detection materials with low atomic numbers, an understanding of the atmospheric neutron environment and its dependencies on time and position is needed. To produce a simple equation based model, Monte Carlo simulations were performed to obtain the atmospheric neutron fluxes produced by charged galactic cosmic ray interactions with the atmosphere. Based on the simulation results the omnidirectional neutron environment was parametrised including dependencies on altitude, magnetic latitude and solar activity. The upward- and downward-moving component of the atmospheric neutron flux are considered separately. The energy spectra calculated using these equations were found to be in good agreement with data from a purpose built balloon-borne neutron detector, high altitude aircraft data and previously published simulation based spectra.

\end{abstract}

\begin{keyword}
Atmospheric neutrons \sep Monte Carlo \sep Cosmic rays \sep Instrumental background


\end{keyword}

\end{frontmatter}



\section{Introduction}

Instruments flown on balloon-borne and Earth-orbiting satellites experience radiation from a wide range of particles with a galactic, solar, magnetospheric or atmospheric origin \cite{Vianio}. The neutron component is mainly produced in cosmic ray induced showers in the Earth's atmosphere. The production energies for neutrons range from $\sim$0.1 MeV to $\sim$10's of GeV. After production the neutrons are thermalised due to scattering interactions with atmospheric nuclei, resulting in an energy spectrum ranging from eV energies up to 10's of GeV. The non-thermal component of this neutron spectrum is an important source of background for experiments making use of detection materials with low atomic numbers, e.g. plastic scintillators. Examples include X-ray polarimeters operating on balloon experiments, see for example \cite{PoGO},\cite{Mark}, in earth orbiting satellites \cite{Polar}, or balloon-borne Compton telescopes \cite{Volker}. A second source of measurement background, found in a wider range of experiments, stems from  neutron-induced activation of both active and passive materials in the detector \cite{astroH}. Reducing the neutron-induced background through active shielding is inefficient due to the non-ionizing behaviour of neutrons, while passive shielding from neutrons results in a significant mass increase of the payload. As a result an irreducible and variable background resulting from neutrons is often unavoidable. During the design phase of the instrument an understanding of the incoming neutron flux is therefore needed to optimise the signal to background of the experiment. 

The spectral shape and the energy integrated flux of atmospheric neutrons varies strongly with altitude. The energy integrated flux has additional dependencies on the magnetic latitude and solar activity. For Earth orbiting satellites and long duration balloon experiments the latitude, solar activity and altitude may not be constant during the mission, resulting in a variable neutron-induced background rate. An understanding of how the neutron flux and spectral shape changes throughout the mission is therefore important. The neutron flux is largest and most influenced by solar activity in the polar regions, where long duration stratospheric balloon flights are conducted \cite{PoGO2}. 

Recently published studies have verified that the atmospheric neutron environment can be simulated accurately using different Monte Carlo packages, see for example \cite{Nesterenok},\cite{Sato},\cite{Overholt},\cite{Kovaltsov},\cite{Masarik}. The Monte Carlo based packages PLANETOCOSMICS \cite{PLANETO} and a combination of Geant4 \cite{Geant4} and MCNP \cite{MCNP} are used respectively in \cite{Kovaltsov} and \cite{Masarik} to simulate the neutron flux dependent cosmogenic nuclei production rates. The neutron production rates are simulated for a more general purpose using Geant4 \cite{Geant4}, PHITS \cite{PHITS} and a combination of CORSIKA \cite{CORSICA}, MCNP \cite{MCNP} and MCNPX \cite{MCNPX} respectively in \cite{Nesterenok}, \cite{Sato} and \cite{Overholt} . In particular the work presented in \cite{Nesterenok} provides a look-up table for neutron spectra at particular altitudes, all cut-off rigidities and all solar activities, which are shown to match spectra as measured on ground and on high altitude flights. In \cite{Sato} analytical functions are furthermore provided describing the neutron energy spectra over a wide energy range for altitudes below $20\,\mathrm{km}$. Also in \cite{Sato} the results are shown to be in good agreement with spectra measured on ground and at flight altitudes.

The aim of this paper is to present a relatively simple model which provides the spectrum of the non-thermal neutron component, for all altitudes exceeding $5\,\mathrm{km}$, which can easily be implemented in, for example, Geant4 based simulations of balloon-borne and Earth-orbiting instruments. Such a model can, for example, be used to study the effect of variations in  position and solar activity on the neutron flux impinging on an astrophysics experiment. For this purpose a Monte Carlo data based set of equations describing the atmospheric neutron flux in the energy range of $8\,\mathrm{keV}$ to $1\,\mathrm{GeV}$ for altitudes exceeding $5\,\mathrm{km}$, all magnetic latitudes and all solar activities is presented. The directional dependency of the neutron flux on altitude and energy is of additional importance, especially for balloon-borne experiments. The direction of momentum is therefore included in the model. Section 2 provides a brief overview of the atmospheric neutron environment. This is followed by a description of the performed Monte Carlo simulations in section 3. Section 4 presents the results of these simulations and the resulting parametrised model and comparisons with results from other models and measurement data. The energy spectrum is divided into an upward and downward moving component in section 5. Finally section 6 discusses the neutron environment specifically for high latitude balloon-borne instruments using a comparison with data from a purpose-built neutron detector.

\section{Atmospheric Neutrons}

The majority of atmospheric neutrons are produced in hadronic air showers induced by cosmic ray protons or helium nuclei \cite{Hess}. A second source is electromagnetic air showers induced either by cosmic gamma rays or electrons. The hadronic component of electromagnetic showers in which neutrons are produced is, however, small. Atmospheric neutrons can also be produced by radioactivity in the Earth. Due to the high atmospheric density at low altitudes, which results in a relatively short mean free path for these neutrons, the contribution from the Earth is only relevant in the lower part of the troposphere. Only production through hadronic air showers is therefore considered here.

Within hadronic air showers the highest energy neutrons are produced by charge-exchange interactions between cosmic ray nuclei and atmospheric nuclei \cite{Hess}. The resulting neutrons will carry approximately the momentum of the incoming protons, and will therefore preferentially move downwards in the atmosphere. The cross section for this process is not relevant at sub-GeV energies, as a result neutrons produced through this mechanism typically have kinetic energies exceeding $1\,\mathrm{GeV}$. 

In hadronic air showers the majority of neutrons are produced in the sub-GeV region through head-on collisions of cosmic rays with atmospheric nucleons. During the collision, a nucleon within the atmospheric nucleus gains momentum and forms an intranuclear cascade \cite{Bert}. In this cascade different particles are created, the energy of which, due to the Pauli Exclusion Principle, must exceed the highest occupied energy level in the nucleus. As a result, the intranuclear cascade results in neutrons with energies ranging from 10's to 100's of MeV. Due to the high energies involved, the majority of the momenta of the produced neutrons are directed towards the Earth's surface.

The intranuclear cascade is followed by the emission of particles from the excited remnant nucleus. In this process, often referred to as evaporation, the nucleus moves to its ground state energy by emitting hadrons and photons. The emission of hadrons, including neutrons, continues until the excitation level drops below 10's to 100's of keV, after which the remaining energy is lost through photon emission \cite{Bert}. The neutrons produced through evaporation have typical energies around $1\,\mathrm{MeV}$. The emission from the nucleus proceeds isotropically within the rest frame of the nucleus. The resulting neutron emission is therefore expected to be more isotropic than those coming from the two previously described processes. 

%

After production, neutrons lose energy by scattering off atmospheric nuclei. The energy loss and cross section increase with the decreasing mass of the atmospheric nucleus the neutron interacts with. The process of energy loss, referred to as thermalisation, continues until the energy of the neutron is equal to that of the average energy of the atmospheric nuclei surrounding it. The momentum vector changes at each scattering interaction. As a result, the direction of the neutrons becomes more isotropic during thermalisation. The density gradient of the atmosphere furthermore has an influence on the momentum vector after many scatterings. As a result of this gradient the majority of the scattered neutrons are expected to move upwards in the upper stratosphere. 


\begin{figure}[!t]
\centering
\includegraphics[width=5.0in]{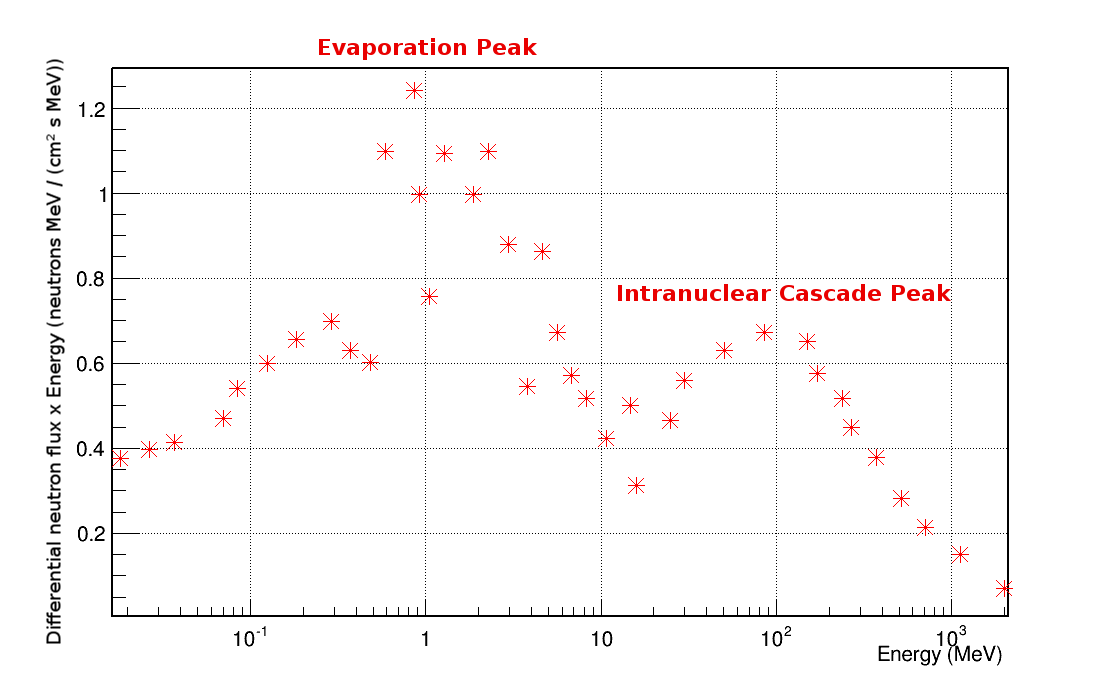}
\caption{A differential neutron energy spectrum multiplied by the neutron energy measured at an altitude of $\sim20\,\mathrm{km}$ and a magnetic latitude of $58^\circ$. The data points were taken from \cite{Goldy}.}
\label{typical}
\end{figure}

A typical differential neutron energy spectrum multiplied by the neutron energy, as measured at an altitude of $\sim20\,\mathrm{km}$ and a magnetic latitude of $58^\circ$, is shown in figure \ref{typical}. By multiplying the spectrum with the energy, the evaporation and intranuclear cascade production energy regions become clearly visible at the respective energies of $\sim1\,\mathrm{MeV}$ and $\sim100\,\mathrm{MeV}$.


The amplitude of the spectrum is not only expected to vary with altitude, but also with magnetic latitude and solar activity. The Earth's magnetic field shields the equator from all charged cosmic rays with a rigidity below $\sim$15 GV, whereas in polar regions the cut-off is below 1 GV. As a result, the neutron flux is highest at the magnetic poles. The solar magnetic field, frozen in the solar wind, further shields the Earth from cosmic rays. This effect can be expressed using the force-field approximation \cite{Axford}. The modulation of the Local Interstellar Spectrum (LIS) by solar activity is approximated by a potential term $\phi$. Typical values of $\phi$, for solar maximum and solar minimum periods, are respectively $1250\,\mathrm{MV}$ and $350\,\mathrm{MV}$ \cite{Usoskin}. The exact value of $\phi$ depends on the model used for the LIS \cite{Usoskin2}. For the work presented here, the proton LIS from \cite{Berger} was used. Furthermore both the solar activity and the Earth's magnetic field have the largest effect on the lowest energy charged cosmic rays. The dependencies from these two magnetic fields are therefore coupled, meaning that the effects from solar activity are most pronounced at the magnetic poles. Due to the relatively small dependency of the proton-air cross section on energy (for protons with energies exceeding $1\,\mathrm{GeV}$) the shape of the differential neutron energy spectrum does not vary significantly with magnetic latitude and solar activity. The spectral shape in the sub-GeV region can therefore be assumed to only vary with altitude. This can be seen in, for example, \cite{Goldy} where measured spectra are presented for different magnetic latitudes and nearly identical altitudes.

\section{Simulations}

In order to simulate the atmospheric neutron environment the PLANETOCOSMICS simulation package \cite{PLANETO}, which incorporated Geant4.9.5.p02 \cite{Geant4} for particle interactions, was used. The QGSP\_BIC\_HP \cite{QGSP} physics list, which uses the Binary Cascade Model to handle protons and neutrons with energies between $20\,\mathrm{MeV}$ and $10\,\mathrm{GeV}$ and the G4NDL4.0 data set \cite{G4NDL}, to handle scattering interactions of neutrons with energies below  $20\,\mathrm{MeV}$, was used. Neutron capture reactions and neutron decay are furthermore taken into account in the simulations. The simulations presented here made use of a spherical model of the Earth (consisting of $\mathrm{SiO_2}$) with a radius of $6371\,\mathrm{km}$. The atmosphere was described using the NRLMSISE00 model \cite{NRM} for $0^\circ$ latitude and longitude and was set to extend up to $100\,\mathrm{km}$. The Earth's magnetic field was described using the IGRF model, for the reference date of January 1st 2000 \cite{IGRF}, for the internal field and the TSY2001 model \cite{TSY} for both the outer magnetic field and the magnetopause. The incoming charged cosmic ray flux comprised only protons and helium nuclei. The fraction consisting of heavier ions, responsible for $1\%$ of the charged cosmic rays, was accounted for by scaling up the proton and alpha spectra. This can potentially lead to an underestimation of the neutron flux.

Cosmic ray protons and alpha particles were generated isotropically from a geocentric spherical shell with a radius of $2\times10^6\,\mathrm{km}$. Neutrons produced in the atmosphere were sampled at altitudes corresponding to pressures of $550\,\mathrm{hPa}\,(\sim$5 km), $234\,\mathrm{hPa}\,(\sim$11.3 km), $100\,\mathrm{hPa}\,(\sim$16 km), $55\,\mathrm{hPa}\,(\sim$20 km), $25\,\mathrm{hPa}\,(\sim$25 km), $12\,\mathrm{hPa}\,(\sim$30 km), $6\,\mathrm{hPa}\,(\sim$35 km), $3\,\mathrm{hPa}\,(\sim$40 km), $0.225\,\mathrm{hPa}\,$ $(\sim$60 km), $0.00025\,\mathrm{hPa}\,(\sim$99 km) and $0\,\mathrm{hPa}$.

Solar modulation effects on the incoming cosmic ray spectra were taken into account by dividing the incoming proton and alpha spectra into separate energy ranges. The incoming cosmic ray spectrum was split up into energy ranges: $0-1.5$, $1.5-2.5$, $2.5-3.5$, $3.5-4.5$, $4.5-5.5$, $5.5-6.5$, $6.5-7.5$, $7.5-8.5$, $8.5-9.5$, $9.5-10.5$ and $10.5-20\,\mathrm{GeV}$. Within these ranges the energy of the incoming cosmic rays was taken to be mono-energetic using the mean energy of the range. The last energy range, between $20-\infty\,\mathrm{GeV}$, was simulated as as a power law with index $-2.7$. Simulations were performed for these individual energy ranges and the resulting spectra were summed using normalisation factors based on cosmic ray spectra, as measured at the top of the atmosphere during different solar activities. Using this method the same simulation data could be used to recreate the neutron environment during different solar activities. The spectra measured by the PAMELA experiment during December 2009 \cite{2009} were used to simulate the neutron environment for a solar minimum ($\phi=250\,\mathrm{MV}$). The neutron environment during a period with high solar activity ($\phi=1109\,\mathrm{MV}$) was simulated using the proton and helium spectra measured by BESS during Summer 2002 \cite{BESS_maximum}.

\section{Results}

\begin{figure}[!t]
\centering
\includegraphics[width=5.0in]{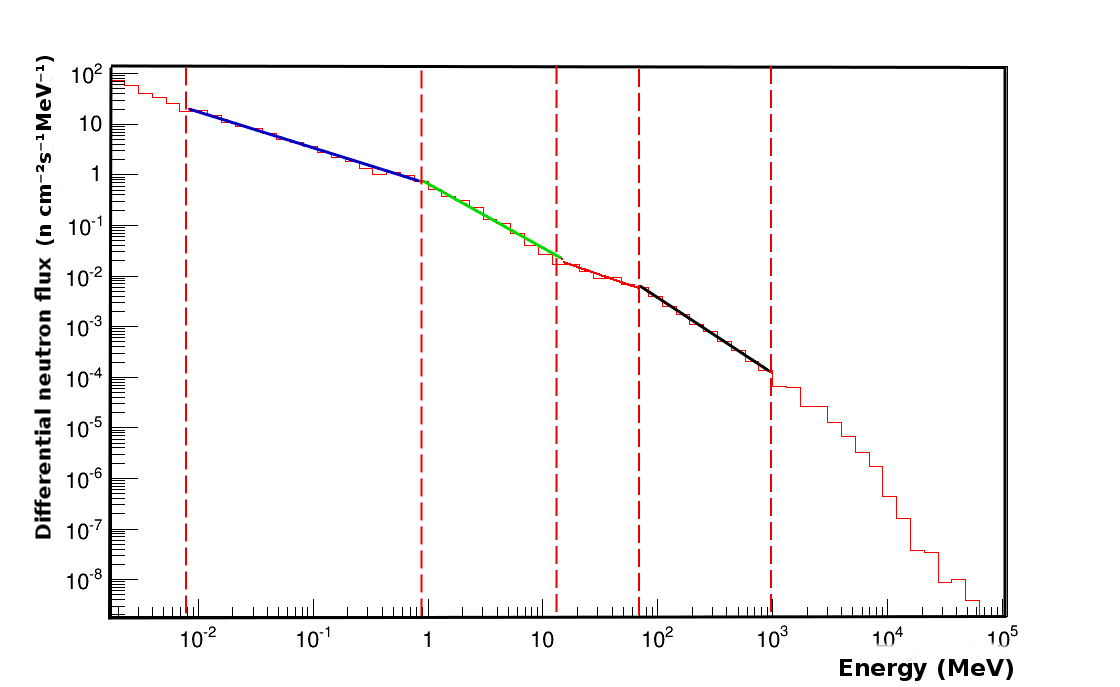}
\caption{An example of the differential neutron flux as a function of energy, simulated using the simulation procedure described in this paper, for an altitude corresponding to $25\,\mathrm{hPa}$ ($\sim$25 km), for a magnetic latitude of $62^\circ$ during a solar minimum. The histogram is fitted using 4 power laws in the ranges, $8-900\,\mathrm{keV}$ (blue), $0.9-15\,\mathrm{MeV}$ (green), $15-70\,\mathrm{MeV}$ (red) and $70-1000\,\mathrm{MeV}$ (black)}
\label{60_64}
\end{figure}

An example of a typical neutron energy spectrum, resulting from the simulations, is shown in figure \ref{60_64}. Below $900\,\mathrm{keV}$ the spectrum is relatively hard and dominated by thermalised neutrons and low energy neutrons produced in the evaporation process. Above the mean production energy for the evaporation process, $\sim0.9\,\mathrm{MeV}$, the flux can be seen to decrease more rapidly with increasing energy. A second kink in the spectrum can be seen around $15\,\mathrm{MeV}$ at the start of the region where the intranuclear cascade becomes the dominant production process. From this point, at $15\,\mathrm{MeV}$, to the mean production of the intranuclear cascade, found at $\sim70\,\mathrm{MeV}$, the slope is less steep. Above $70\,\mathrm{MeV}$ the neutron flux can be seen to drop off sharply as only charge exchange remains as a production process. Based on this shape of the spectrum the simulated neutron differential energy spectra were fitted with 4 power laws of the form $\mathrm{F(neutrons/(cm^2\,s\,MeV))} = \mathrm{A\,E^{-\alpha}}$, where the parameters $A$ and $\alpha$ will be respectively referred to as the normalisation and slope from here on. The best fit results were consistently acquired with the fit ranges of $8\,\mathrm{keV}-0.9\,\mathrm{MeV}$, $0.9\,\mathrm{MeV}-15\,\mathrm{MeV}$, $15\,\mathrm{MeV}-70\,\mathrm{MeV}$ and $70\,\mathrm{MeV}-1000\,\mathrm{MeV}$. The normalisation for the different power laws are referred to as $A,B,C$ and $D$ for the ranges $8-900\,\mathrm{keV}$, $0.9-15\,\mathrm{MeV}$, $15-70\,\mathrm{MeV}$, $70-1000\,\mathrm{MeV}$ respectively. The slopes for these regions will be referred to as $\alpha$, $\beta$, $\gamma$ and $\delta$, respectively. A change in slope at energies below $8\,\mathrm{keV}$, the exact position of which was found to depend both on altitude and on the used evaporation model, defined the lower energy limit. The upper limit of 1 GeV was chosen because of low statistics above this energy and a significant dependency of the spectral shape on magnetic latitude for energies exceeding $1\,\mathrm{GeV}$. 

These fits were performed to all the spectra resulting from the simulations for altitudes above $550\,\mathrm{hPa}$ ($\sim5\,\mathrm{km}$). This lower altitude limit is based on limited statistics and the effect of the Earth's surface composition on the neutron flux at low altitudes. An attempt was made to parametrise the dependency of the fit results on altitude, magnetic latitude and solar activity. 

\subsection{Altitude Dependence}

\begin{figure}[!t]
\centering
\includegraphics[width=5.0in]{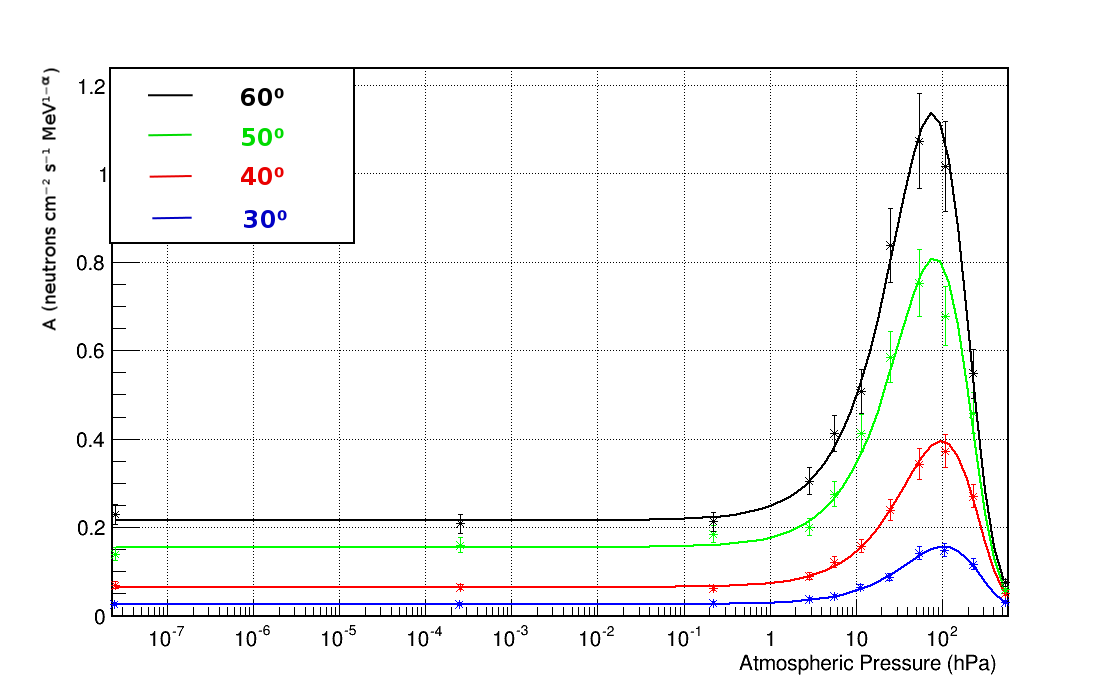}
\caption{The altitude dependence of the normalisation parameter A for 4 different magnetic latitudes ($30^\circ$ in blue, $40^\circ$ in red, $50^\circ$ in green and $60^\circ$ in black) together with the fitted functions of the form $A = \left[ah + b\right]e^{-h/c}+d$ for a solar minimum. The presented error bars are the fitting errors of the power law fits.}
\label{A_vs_alt}
\end{figure}

First the dependency of the fit parameter $A$ on the altitude was studied. The values of $A$, resulting from fitting the Monte Carlo data, were plotted as a function of altitude, while expressing the altitude in atmospheric pressure $h$ (hPa). Figure \ref{A_vs_alt} shows the values of $A$ for the different altitudes and for the magnetic latitudes $30^\circ,\,40^\circ,\,50^\circ$ and $60^\circ$. The simulation results shown in figure \ref{A_vs_alt} are for solar minimum conditions. 

The values of $A$ were subsequently fitted using a function of the form:

\[A = \left[ah+ b\right]e^{-h/c} + d\]

The function contains an exponential decay at high atmospheric pressures described by the term $be^{-h/c}$. At low atmospheric pressures the function contains an exponential increase, described by the term $ahe^{-h/c}$, starting from a plateau level defined by the parameter $d$. The pressure where the exponential decay starts, which coincides with the altitude where the maximum flux is found, is dictated by the parameter $c$. A minimum of 4 parameters is therefore required to accurately describe the altitude dependence of $A$. The reduced $\chi^2$ values ($\chi^2$ divided by the number of degrees of freedom), with 7 degrees of freedom, for the fits of $A$ as a function of energy were found to be in the range of $0.4$ to $1.8$ with a mean value of $0.9$. Similar results were found for solar maximum conditions. It can therefore be concluded that this function describes the altitude dependence of $A$ well for all magnetic latitudes and solar activities. 

\subsection{Magnetic Latitude and Solar Dependence}

\begin{figure}[!t]
\centering
\includegraphics[width=5.0in]{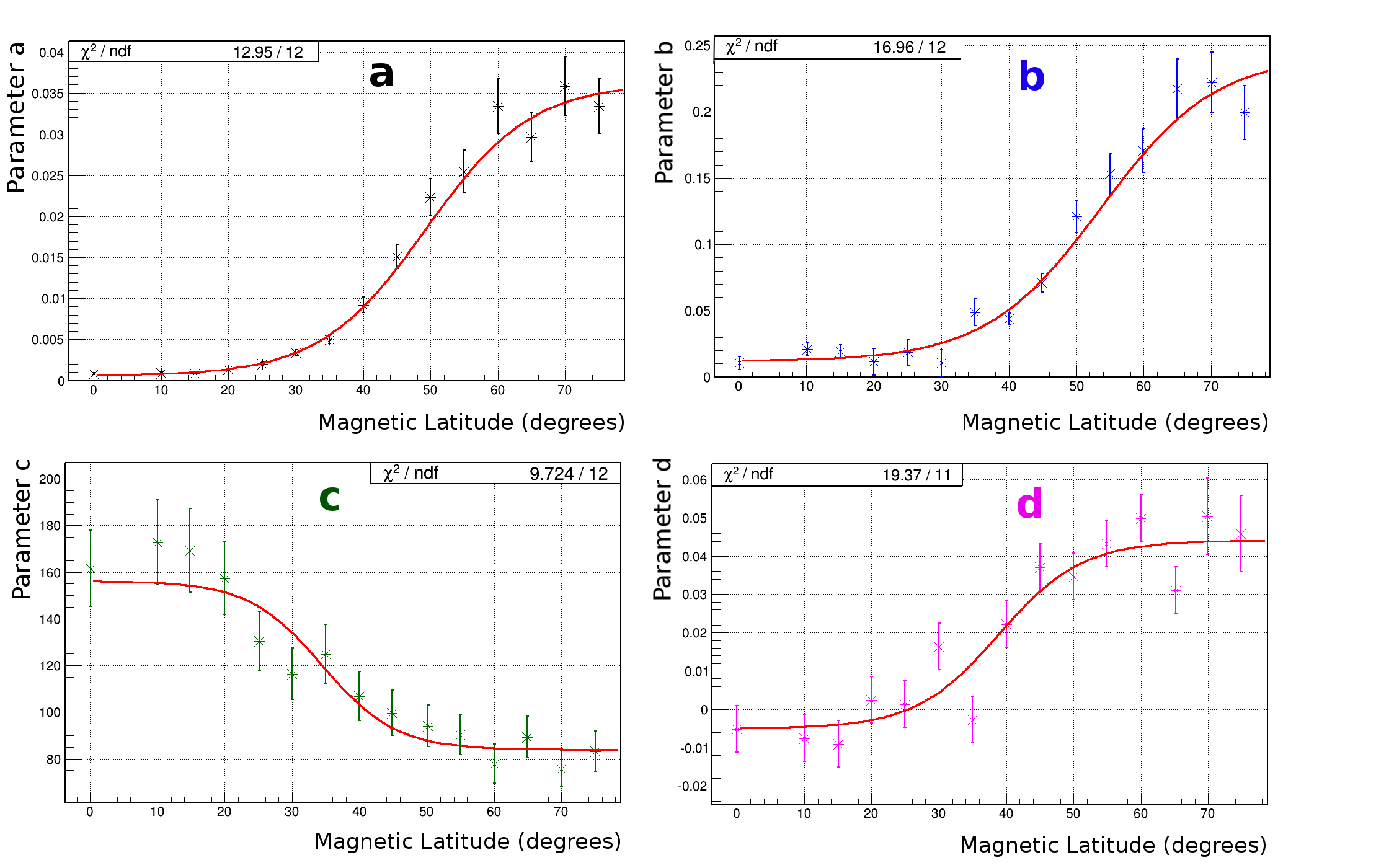}
\caption{Parameters $a$,$b$,$c$ and $d$ plotted as a function of magnetic latitude, for a solar minimum conditions, fitted with functions of the form $p_0+p_1[1-\tanh(p_2x)]$. The presented error bars are the fitting errors.}
\label{a_b_c_d}
\end{figure}

As can be seen from figure \ref{A_vs_alt}, the parameters describing the altitude dependence of $A$ vary with magnetic latitude. They furthermore have a dependency on solar activity. An example of values of $a$,$b$,$c$ and $d$ resulting from the performed fits for parameter $A$, together with the fitting errors are shown in figure \ref{a_b_c_d} as a function of magnetic latitude. All parameters can be seen to vary with magnetic latitude in a similar way which can be described using a $p_0+p_1(1-\tanh(p_2\lambda))$ relationship. A minimum of 3 parameters is required to describe this function accurately, one describing the plateau value, $p_0$, one the amplitude, $p_1$ and one to describe how fast the function rises, $p_2$. The different parameters were fitted using functions of the form $p_0+p_1(1-\tanh(p_2\lambda))$. The reduced $\chi^2$ values for all the performed fits of this type were in the range of $0.5$ and $1.9$, with 11 degrees of the freedom. The average reduced $\chi^2$ was $1.1$. When varying the solar activity the amplitude of the functions was found to vary for parameters $a$,$b$ and $d$. A fourth parameter is therefore required in these functions to describe the solar activity dependence. The parameters $a,b,c$ and $d$ were finally found to be best described as a function of magnetic latitude and solar activity using:

%
%
%

\[a = 6.0\times10^{-4} + (1.85 - 1.35\,\mathrm{S})\times10^{-2} \left[1 - \tanh(180 - 3.5 \mathrm{\lambda})\right]  \]

\[b = 1.1\times10^{-2} + (1.2 - 0.8\,\mathrm{S})\times10^{-1} \left[1 - \tanh(180 - 3.5 \mathrm{\lambda})\right]  \]

\[c = 150 - 33 \left[1 - \tanh(180 - 5.5 \mathrm{\lambda})\right]  \]

\[d = -4.0\times10^{-3} + (2.4 - 1.0\,\mathrm{S})\times10^{-2} \left[1 - \tanh(180 - 4.4 \mathrm{\lambda})\right]  \]

Where $\mathrm{\lambda}$ is the magnetic latitude, in degrees, and $\mathrm{S}$ the solar activity parameter. The value of S is zero for solar minimum with $\phi=250\,\mathrm{MV}$ and is equal to unity at a solar maximum with $\phi=1109\,\mathrm{MV}$. These values were chosen to match the solar activity persisting during the measurements of the PAMELA and BESS instruments during solar minimum and solar maximum conditions respectively. S furthermore varies linearly with $\phi$. It should be noted here that the incoming cosmic ray spectrum for a given magnetic latitude varies with geomagnetic conditions. The parametrisation is therefore only valid for standard geomagnetic conditions similar to those on January 1st 2000 and may be different for other periods of time.

\subsection{Parameter $A$}

\begin{figure}[!t]
\centering
\includegraphics[width=5.0in]{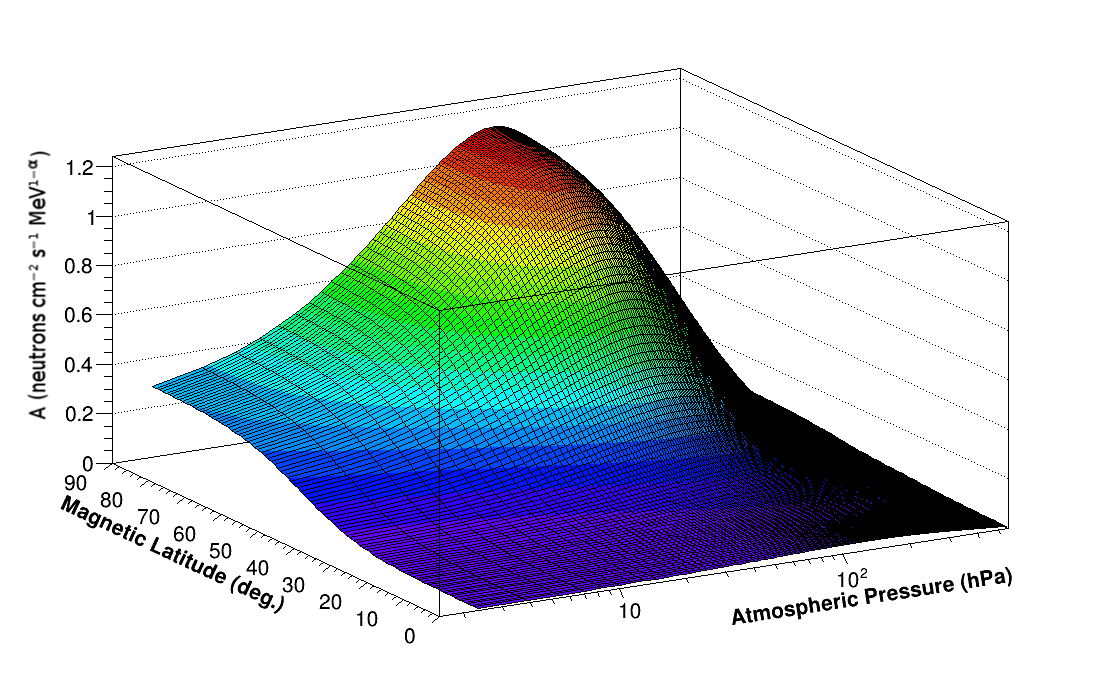}
\caption{The normalisation parameter A as a function of altitude and magnetic latitude for a solar minimum.}
\label{A_lat_lat}
\end{figure}

Using the dependencies of $A$ found through the fitting procedure, the value of $A$ can be plotted as a function of altitude and latitude. In figure \ref{A_lat_lat} this is done for solar minimum conditions. For magnetic latitudes corresponding to the cut-off rigidities below $\sim1\,\mathrm{GV}$ the dependency of $A$ on magnetic latitude can be seen to be small. This is a result of the decreasing neutron production cross section for protons with energies below $1\,\mathrm{GeV}$. The dependency of A on magnetic latitude can be seen to be large at the mid-latitude regions where the increase in cosmic ray flux, with energies above $1\,\mathrm{GeV}$, with increasing magnetic latitude is large. At lower magnetic latitudes the dependency of $A$ on magnetic latitude is less strong due to the smaller increase of incoming cosmic ray flux with increasing magnetic latitude in this region. At low magnetic latitudes the cut-off rigidity is furthermore not constant within a set magnetic latitude range, for example, the cut-off rigidity varies by several GV at the magnetic equator. The value of $A$ given here is averaged for a given magnetic latitude, this effect further reduces the dependency of $A$ on the magnetic latitude in this region. It should be stressed that at these low magnetic latitudes a further dependency on magnetic longitude also exists but is not taken into account in this model. Potential effects induced by the South Atlantic Anomaly are furthermore not accounted for using this approach.

\begin{figure}[!t]
\centering
\includegraphics[width=5.0in]{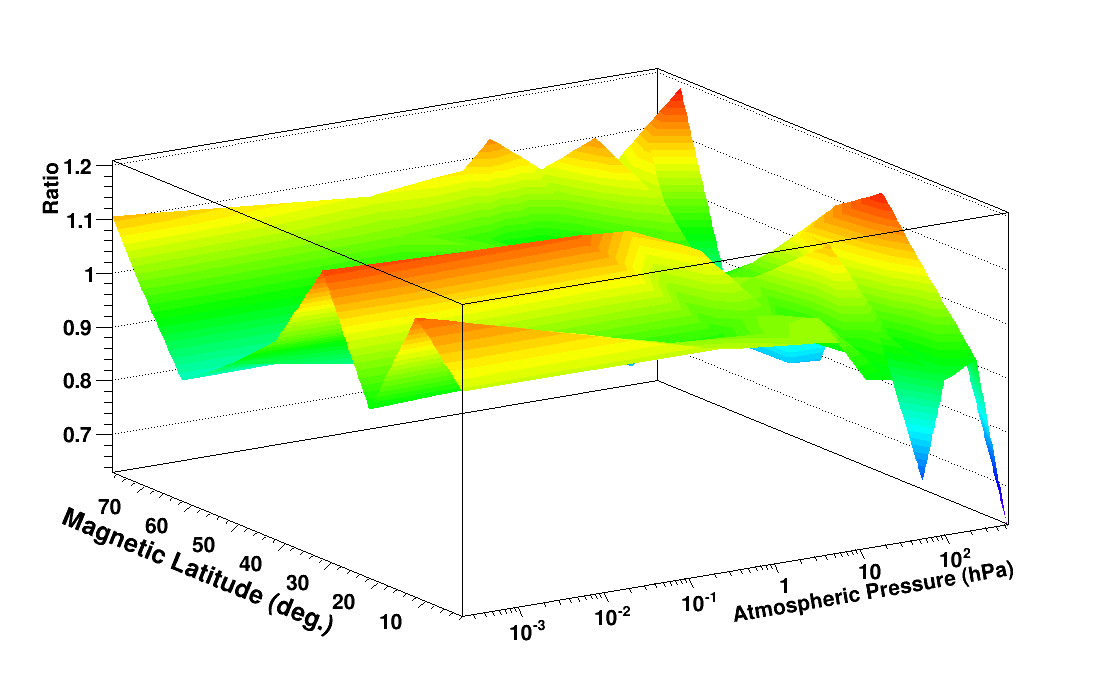}
\caption{The ratio of parameter $A$ as resulting from the model and the Monte Carlo data. Both the values of the model and the Monte Carlo data are for solar minimum conditions. The ratio values are distributed as a Gaussian around 1 and have a standard deviation of $9\%$.}
\label{resid_min}
\end{figure}

\begin{figure}[!t]
\centering
\includegraphics[width=5.0in]{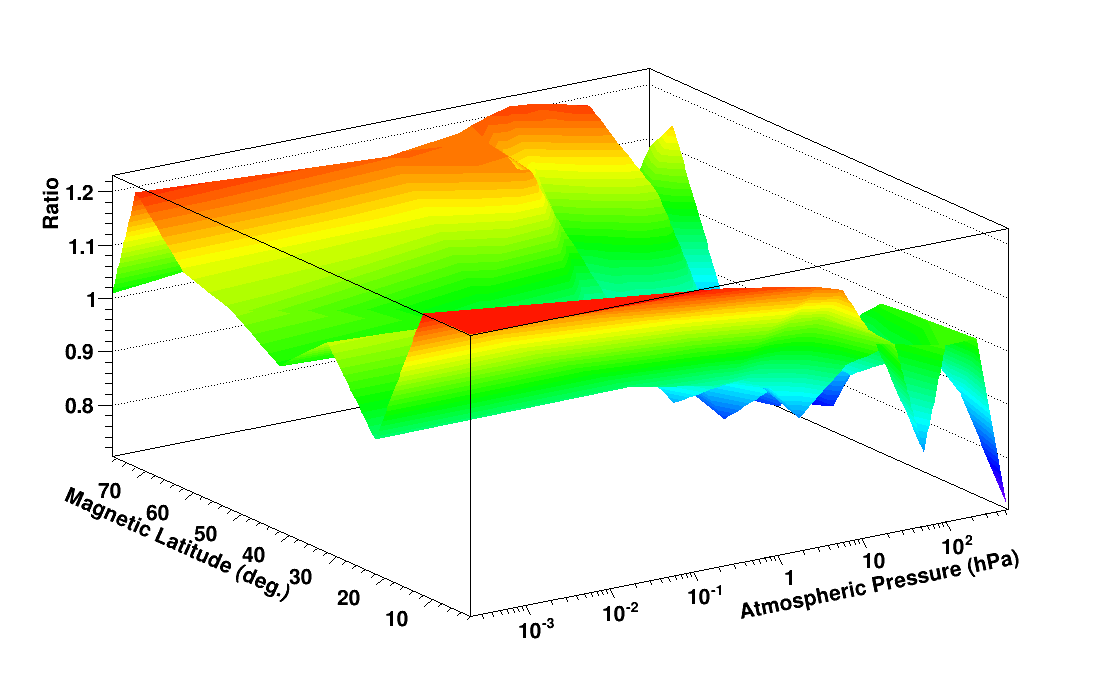}
\caption{The ratio of parameter $A$ as resulting from the model and the Monte Carlo data. Both the values of the model and the Monte Carlo data are for solar maximum conditions. The ratio values are distributed as a Gaussian around 1 and have a standard deviation of $13\%$.}
\label{resid_max}
\end{figure}

The deviations in values of $A$, resulting from the model, from the original Monte Carlo data are shown in figures \ref{resid_min} and \ref{resid_max}, for solar minimum and solar maximum conditions respectively. The largest deviations are observed in altitudes and latitudes with small values of $A$. At these locations the Monte Carlo data is statistically limited. The values as shown in figures \ref{resid_min} and \ref{resid_max} are distributed as a Gaussian with a mean value $1.0$ and a standard deviation of $0.09$ and $0.13$ respectively, indicating a relative difference of $9\%$ and $13\%$ between the values of $A$ resulting from this parametrisation model and the original Monte Carlo data for these two situations. The average error on $A$ as a result of the fitting procedure is therefore assumed to be $11\%$.

\subsection{Slopes}

After modelling the normalisation parameter $A$, a similar procedure was used to find the dependency of the slope parameters on altitude. By fitting the Monte Carlo data the slope parameters $\alpha$, $\beta$ and $\delta$ were found to be best approximated using exponentials reaching a constant value at high atmospheric pressures. A minimum of three parameters is therefore required to describe the dependencies of the slopes on altitude. The parameter $\gamma$ was additionally found to decrease exponentially at high atmospheric pressures, an extra parameter was therefore required to best represent this dependency on altitude. The functions, together with their uncertainties, found through fitting the Monte Carlo data, to describe the altitude dependence of the different slopes are:

\[\alpha = -(0.281\pm0.003)\,e^{-h/(4.6\pm0.1)} + (0.732\pm0.002)  \]

\[\beta = -(0.186\pm0.005)\,e^{-h/(13.2\pm1.2)} + (1.308\pm0.003)  \]

\[\gamma = \left[(0.011\pm0.001)\,h + (0.30\pm0.03)\right]\,e^{-h/(68.0\pm9.7)} + (0.26\pm0.03)  \]

\[\delta =  (0.66\pm0.03)\,e^{-h/8.5\pm0.6} + (1.40\pm0.02)  \]

\begin{figure}[!t]
\centering
\includegraphics[width=5.0in]{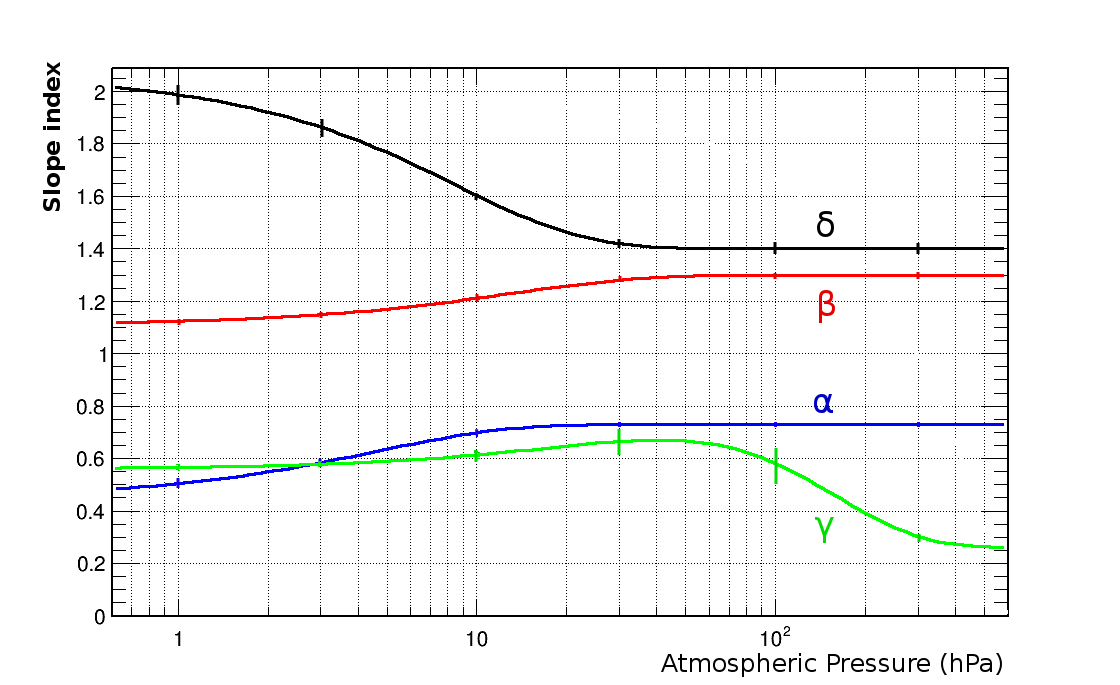}
\caption{The slope parameters $\alpha$ (blue), $\beta$ (red), $\gamma$ (green) and $\delta$ (black) as a function of atmospheric pressure (hPa). The uncertainties, resulting from the fitting errors, are shown at pressures: 1,3,10,30,100 and $300\,\mathrm{hPa}$.}
\label{slopes}
\end{figure}

Here $h$, in hPa, is again used to express the altitude. The parameters $\alpha$, $\beta$, $\gamma$ and $\delta$ resulting from this parametrisation, are shown as a function of altitude in figure \ref{slopes} together with their uncertainties. No significant dependencies on magnetic latitude and solar activity were found for the slope parameters. Due to the smaller number of dependencies the fitting errors are relatively small with respect to those on parameter $A$.

Using the combination of $\mathrm{A},\, \alpha\,, \beta \,, \gamma$ and $\delta$, the normalisations $B,\, C$ and $D$ can be calculated using:

\[B = A\,0.9^{-\alpha+\beta}\]

\[C = B\,15^{-\beta+\gamma}\]

\[D = C\,70^{-\gamma+\delta}\]

Using the 4 different normalisation parameters and the 4 slope parameters the atmospheric neutron spectra in the energy range $8\,\mathrm{keV}-1\,\mathrm{GeV}$ can now be calculated. It should be stressed that this parametrisation has no direct relationship to physical processes; rather, they were chosen to well describe the Monte Carlo data. In the following subsection the results of this parametrisation will be compared to other works.

\subsection{Comparisons to Other Work}

\begin{figure}[!t]
\centering
\includegraphics[width=5.0in]{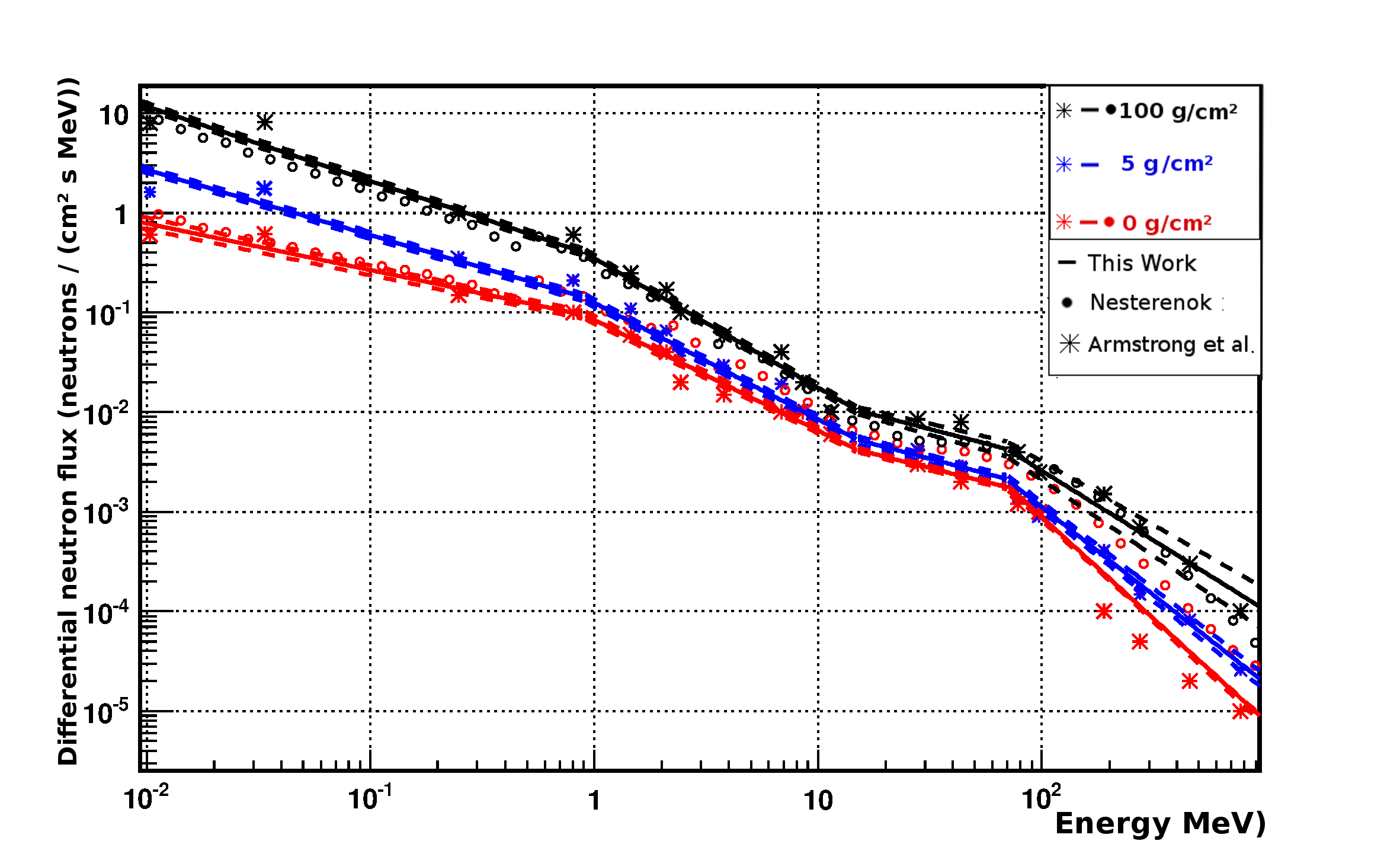}
\caption{Simulation results of the differential neutron flux for three different altitudes as extracted from \cite{Armstrong} (stars), \cite{Nesterenok} (circles) and as predicted by  the model presented in this paper (solid lines). The uncertainties from the model presented here are indicated by the dashed lines. The atmospheric overburden of $0\,\mathrm{g/cm^2}$  is shown in red. The atmospheric overburden of $5\,\mathrm{g/cm^2}$ is shown in blue. The atmospheric overburden of $100\,\mathrm{gr/cm^2}$  is shown in black. For the results from \cite{Armstrong} the incoming cosmic ray spectrum was assumed to be a perfect power law, results from the model presented here are calculated using $\phi = 250\,\mathrm{MV}$, a magnetic latitude of $42^\circ$. For the simulation results from \cite{Nesterenok} a cut-off rigidity of $3.8\,\mathrm{GV}$ is used. }
\label{armstrong_comp}
\end{figure}

The neutron spectra predicted by this model can be compared to previously published location and time specific Monte Carlo simulations \cite{Armstrong} which have been used in the astrophysics community as input for neutron background simulations, see for example \cite{PoGO}, \cite{astroH}. In \cite{Armstrong} an unmodulated primary cosmic-ray spectrum with a cut-off energy at $3.8\,\mathrm{GeV}$ for protons and $6.3\,\mathrm{GeV}$ for helium nuclei was used. The results from \cite{Armstrong} for three different altitudes, corresponding to $0\,\mathrm{hPa}$, $5\,\mathrm{hPa}$ and $98\,\mathrm{hPa} (=100\,\mathrm{g/cm^2})$, were compared with neutron spectra for a magnetic latitude of $42^\circ$ for extreme solar minimum conditions ($\phi = 250\,\mathrm{MV}$) resulting from the parametrisation model presented in this work. The results are furthermore compared to spectra for the altitudes corresponding to $0\,\mathrm{g/cm^2}$ and $\sim100\,\mathrm{g/cm^2}$ from \cite{Nesterenok} for the same magnetic latitude and solar activity (results for $100\,\mathrm{g/cm^2}$ are not provided in \cite{Nesterenok}, $105\,\mathrm{g/cm^2}$ was therefore chosen). The results of the comparison are shown in figure \ref{armstrong_comp}. The spectra can be seen to be in relatively good agreement for the three altitudes considered. Spectra for the altitude corresponding to $5\,\mathrm{g/cm^2}$ are not provided in \cite{Nesterenok}.

\begin{figure}[!t]
\centering
\includegraphics[width=5.0in]{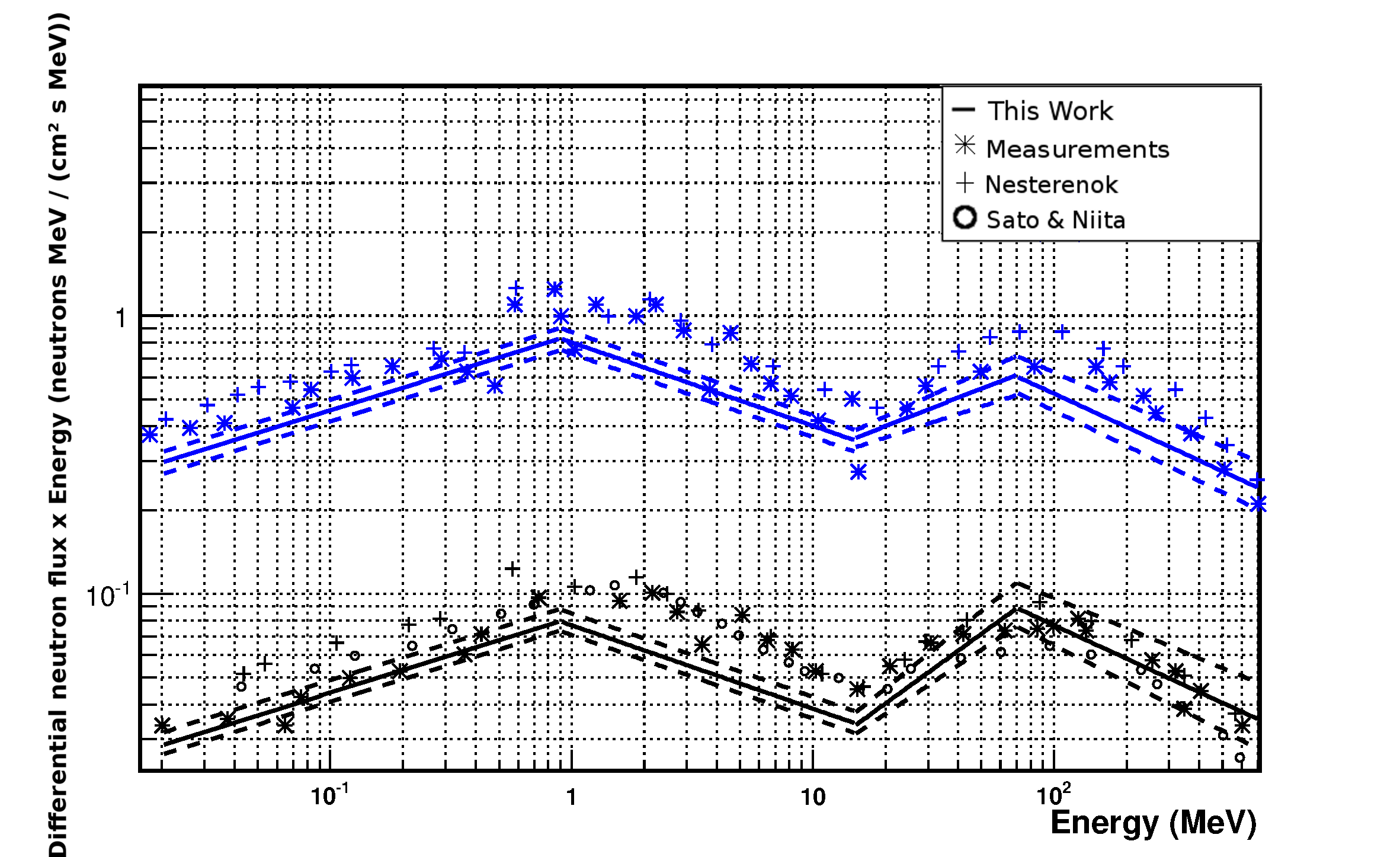}
\caption{The differential neutron flux multiplied by the energy from measurement data extracted from \cite{Goldy} (blue) and from measurement data extracted from \cite{Naka} (black). The data from \cite{Goldy} (star) was taken on an aircraft at an altitude of $20\,\mathrm{km}$ at $54^\circ\,\mathrm{N}$,  $117^\circ\,\mathrm{W}$, during the summer of 1997. The data from \cite{Naka} was taken on a plane at an altitude of $11.28\,\mathrm{km}$ on February 27, 1985. The data from \cite{Goldy} is compared to spectra from \cite{Nesterenok} (cross) and as predicted by the presented model (blue line) for an altitude of 20 km (pressure of $55\,\mathrm{hPa}$ or atmospheric overburden of $56\,\mathrm{g/cm^2}$), magnetic latitude of $60.2^\circ$ and $\phi= 405\,\mathrm{MV}$. The data from \cite{Naka} (star) is compared to spectra as predicted by \cite{Nesterenok} (cross) and \cite{Sato} (circle) and the presented model (line) for an altitude of 11.28 km (pressure of $214\,\mathrm{hPa}$), magnetic latitude of $26.0^\circ$ and $\phi= 656\,\mathrm{MV}$. The uncertainty of the presented model is indicated by the dashed lines. The data points were presented without errors in \cite{Goldy} and \cite{Naka}. }
\label{Goldy_Naka_me}
\end{figure}

There is a paucity of data from stratospheric balloons. Further comparisons therefore focus on data collected by high altitude aircraft for two latitudes which were previously compared to simulation results presented in \cite{Nesterenok} and \cite{Sato}. Model predictions are compared to measured spectra from \cite{Goldy} for an altitude of $20\,\mathrm{km}$, a magnetic latitude of $60.2^\circ$ and $\phi = 405\,\mathrm{MV}$. The solar activity parameter was obtained from \cite{Usoskin}. The spectrum as predicted by the model (for a pressure of $55\,\mathrm{hPa} (=56\,\mathrm{g/cm^2})$) is shown together with the data and the simulation results from \cite{Nesterenok} in figure \ref{Goldy_Naka_me}. Relatively good agreement is found. The largest discrepancy with the data is found in the MeV region, where the measured data contains several spikes and dips which are not present in the spectra resulting from the work presented here, since here the spectrum is assumed to be a simple broken power law. The largest discrepancy with the results from \cite{Nesterenok} is found in the region around $100\,\mathrm{MeV}$. A second comparison, using data from \cite{Naka} taken at a latitude of $26^\circ$, at an altitude of $11.28\,\mathrm{km}$ and with $\phi=656\,\mathrm{MV}$ (the solar activity parameter was again obtained from \cite{Usoskin}) is further shown in figure \ref{Goldy_Naka_me}. For these conditions a comparison to the analytical function derived in \cite{Sato} is presented. 

\section{Directional dependence}

For Earth-orbiting instruments, all atmospheric neutrons can be assumed to enter the satellite from below. For balloon-borne instruments, a non-negligible fraction of neutrons will impinge from above. The ratio of downward to upward moving neutrons will depend strongly on the altitude of the balloon and on the energy of the neutrons. The omnidirectional spectrum was therefore divided into an upward and a downward moving component. The energy spectra for both cases were parametrised using the method described previously. 

\subsection{Upward component}

For the upward-moving component, the following set of equations was found:

\[A = \left[ah+b\right]e^{-h/c}+d  \]

where $a,b,c$ and $d$ vary with magnetic latitude $\lambda$ and solar activity parameter $\mathrm{S}$ according to:

\[\mathrm{a} = 3.0\times10^{-4} + (7.0 - 5.0\,\mathrm{S})\times10^{-3}\left[1 - \tanh(180 - 4.0 \mathrm{\lambda})\right]  \]

\[\mathrm{b} = 1.4\times10^{-2} + (1.4 - 0.9\,\mathrm{S})\times10^{-1} \left[1 - \tanh(180 - 3.5 \mathrm{\lambda})\right]  \]

\[\mathrm{c} = 180 - 42 \left[1 - \tanh(180 - 5.5 \mathrm{\lambda})\right]  \]

\[\mathrm{d} = -8.0\times10^{-3} + (6.0 - 1.0\,\mathrm{S})\times10^{-3} \left[1 - \tanh(180 - 4.4 \mathrm{\lambda})\right]  \]

A behaviour similar to that of the omnidirectional spectrum can be observed. The slope parameters were found to be best described using:

\[\alpha = -(0.290\pm0.005)e^{-h/(7.5\pm0.4)} + (0.735\pm0.004)  \]

\[\beta = -(0.247\pm0.008)e^{-h/(36.5\pm5)} + (1.40\pm0.00)  \] 

\[\gamma = -(0.40\pm0.05)e^{-h/(40\pm10)} + (0.90\pm0.05)  \]

\[\delta =  -(0.46\pm0.03)e^{-h/(100\pm11)} + (2.53\pm0.03)  \] 

\subsection{Downward component}

For the downward-moving component, the following set of equations was found to best describe simulated results:

\[A = \left[ah-b\right]e^{-h/c}+b  \]

Note here that parameter $d$, as used in the omnidirectional and the upward moving flux models, is missing and set equal to $b$. As a result $A$ will tend to zero at high altitudes, representing the vanishing downward moving flux. For the downward component, $a,b$ and $c$ were found to vary with magnetic latitude and solar activity according to:

\[\mathrm{a} = 3.0\times10^{-4} + (1.1 - 0.8\,\mathrm{S})\times10^{-2} \left[1 - \tanh(180 - 3.5 \mathrm{\lambda})\right]  \]

\[\mathrm{b} = 1\times10^{-3} + (1.5 - 0.75\,\mathrm{S})\times10^{-2}\left[1 - \tanh(180 - 4.0 \mathrm{\lambda})\right]  \]

\[\mathrm{c} = 1.40\times10^{2} - 33 \left[1 - \tanh(180 - 5.0 \mathrm{\lambda})\right]  \]

The slope parameters were found to be best approximated using:

\[\alpha = 0.738\pm0.003  \]

\[\beta = 1.270\pm0.003  \]

\[\gamma = ((0.007\pm0.001)h+(0.84\pm0.02))e^{-h/(52\pm5)} + (0.110\pm0.005)  \]

\[\delta =  -(0.27\pm0.03) e^{-h/(230\pm40)} + (1.45\pm0.02)  \]

\begin{figure}
        \centering
        \begin{subfigure}[b]{0.8\textwidth}
                \includegraphics[width=\textwidth]{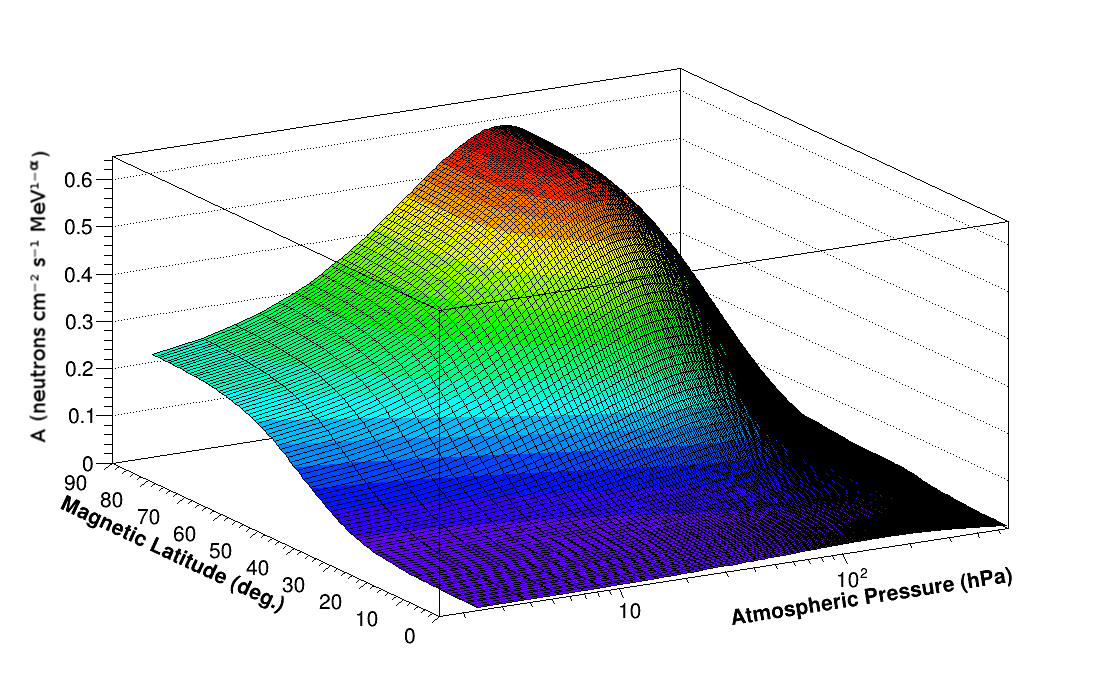}
                \caption{Normalisation parameter A for the upward moving spectrum as a function atmospheric pressure and magnetic latitude for a solar minimum ($\phi=250\,\mathrm{MV}$).}
                \label{gull}
        \end{subfigure}%
        \\
        \begin{subfigure}[b]{0.8\textwidth}
                \includegraphics[width=\textwidth]{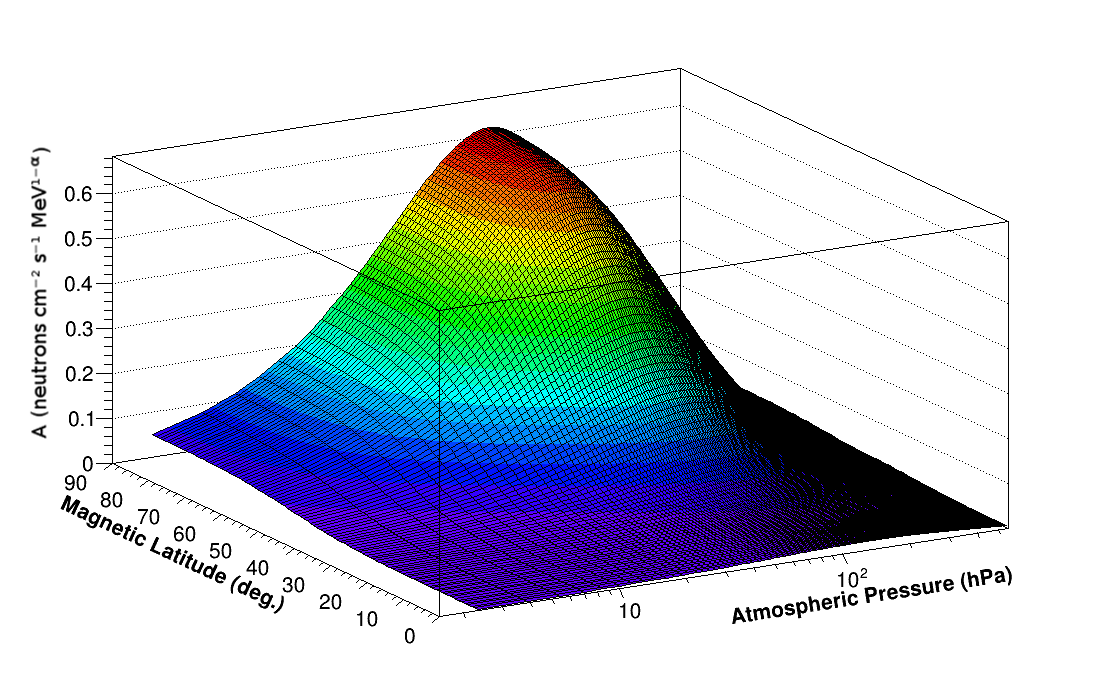}
                \caption{Normalisation parameter A for the downward moving spectrum as a function atmospheric pressure and magnetic latitude for a solar minimum ($\phi=250\,\mathrm{MV}$).}
                \label{tiger}
        \end{subfigure}
        \caption{}\label{animals}
\end{figure}

\begin{figure}
 \includegraphics[width=5.0in]{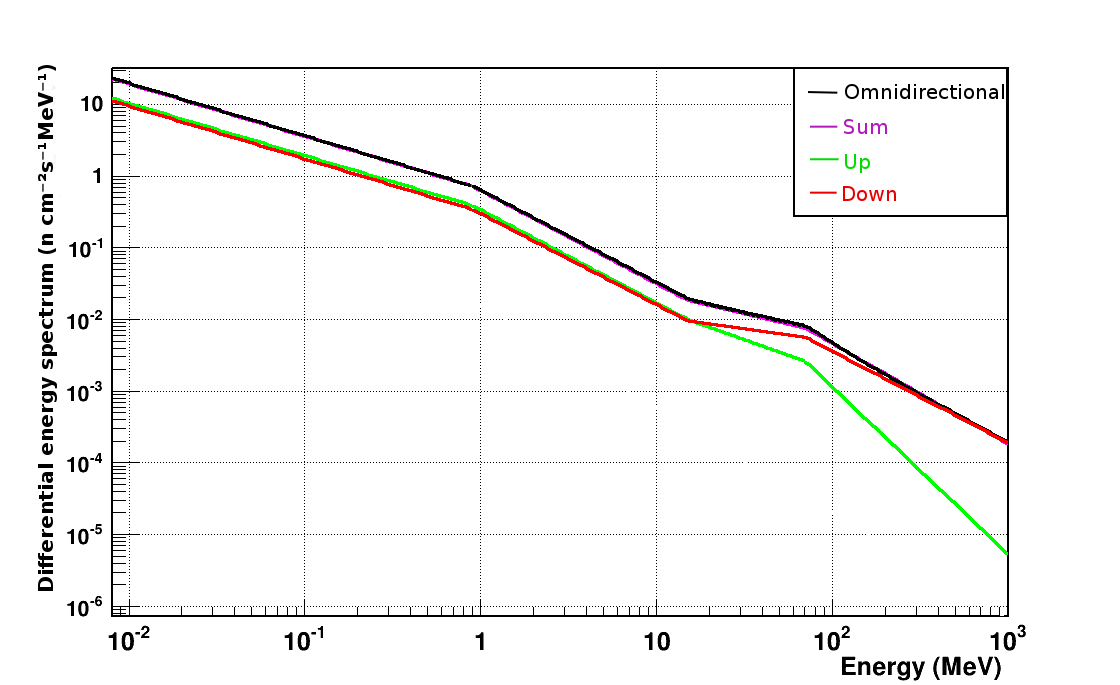}
  \caption{An example of the upward (green) and downward component (red) of the differential energy spectra for an altitude equivalent to $10\,\mathrm{hPa}$, magnetic latitude of $62^\circ$ and a solar minimum ($\phi=250\,\mathrm{MV}$) together with the total spectrum as predicted by the model for the same position. The sum of the upward downward component is shown in magenta and is almost fully covered by the omnidirectional spectrum.}
  \label{mouse}
\end{figure}

Figure \ref{animals} shows the normalisation parameter, A, both for the upward (figure \ref{gull}) and downward component (figure \ref{tiger}). Similar to the omnidirectional case the standard deviations of the relative differences between the fitted parameters $A$ and the Monte Carlo data were calculated. For the upward and downward component respective deviations of $14\%$ and $16\%$ were found. These deviations are larger than the omnidirectional case as a result of lower statistics resulting from dividing the Monte Carlo data in an upward and downward component. The reproduced spectra for an altitude of 35 km, magnetic latitude of 62 degrees during a solar minimum are shown in figure \ref{mouse}. It can be seen that the sum of the upward and downward component matches the omnidirectional spectrum over the full energy range.

\section{Predictions for high latitude balloon flights}

\begin{figure}[!t]
\centering
\includegraphics[width=5.0in]{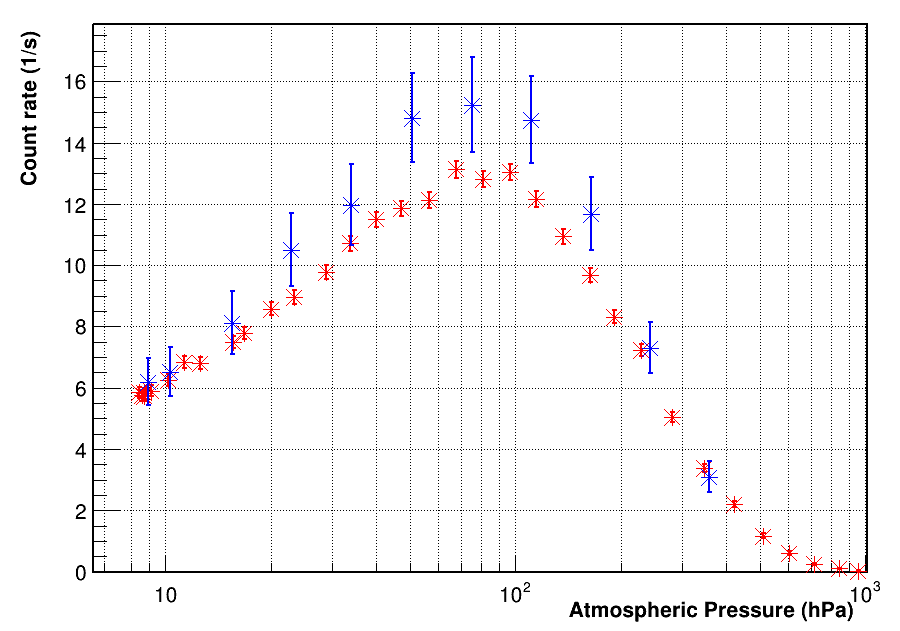}
\caption{The measured neutron flux \cite{myself} as a function of altitude (red) compared to simulation results derived using the model described in his paper for a magnetic latitude of $65^\circ$ and $\phi=800\,\mathrm{MV}$ (blue).}
\label{pogolino}
\end{figure}

The predictions from this direction dependent model have been tested against data from a neutron detector flown on a stratospheric balloon during March 2013 from the Esrange Space Centre in Northern Sweden \cite{myself}. Neutron detection was performed by the instrument using europium doped LiCAF scintillator crystals \cite{Eu} sandwiched between 2 BGO crystals serving as an anti-coincidence system. The neutron detection efficiency of the instrument is approximately constant as a function of neutron energy in the range of 1 eV to 1 MeV, above which the detection efficiency drops off steeply \cite{myself}. The instrument performed data taking during ascent up to a float altitude of 31 km where additional data was collected for approximately one hour before the flight was terminated. The neutron spectra, calculated for different altitudes and a latitude of 65 degrees using the presented model, were used as input in Geant4 simulations of the detector and resulted in the counting rates shown in figure \ref{pogolino}. The solar activity during the flight was approximately $\phi=800\,\mathrm{MV}$. The relatively high value of $\phi$ is a result of a significant Forbush decrease during the flight period. Geomagnetic conditions were however stable. The value of $\phi$ was acquired using data from the Oulu Neutron Monitor \cite{Oulu}, located at a magnetic latitude of $62^\circ$,  and the results presented in \cite{Usoskin}. The detector simulation is described in detail elsewhere \cite{myself}. The simulated results for the detector can be seen to be in relatively good agreement with the measured counting rates. A potential source of discrepancy is the simplification of dividing the incoming flux only into an upwards and downward component. The effect of this is most prominent at the altitudes where most neutrons are produced, between $15$ and $20\,\mathrm{km}$. The relative error from the model at these altitudes is $\sim15\%$. Implementation of a more detailed angular dependency of the neutron flux is expected to result in a better agreement with measurements. Such an implementation would however result in a significant increase in the number of required parameters in the model.

\section{Conclusion}

Simulations were performed using a Geant4 based simulation package, PLANETOCOSMICS, to provide atmospheric neutron spectra in the keV to GeV energy range for different altitudes, latitudes and solar activities. The results from these simulations were used to parametrise the neutron energy spectra in the energy range from 8 keV to 1 GeV for altitudes above 5 km. The results can be used to study fluctuations of the neutron environment resulting from variations in position and solar activity. The spectra calculated using this model were compared to the results presented in \cite{Armstrong}, \cite{Nesterenok} and \cite{Sato} for different altitudes an were found to be in good agreement. Further comparisons with data, taken at largely differing latitudes and altitudes, as presented in \cite{Goldy}, \cite{Naka} and \cite{myself} were also found to be good agreement. It can therefore be concluded that the simple parametrisation model presented here can be used to accurately predict the neutron environment and its variations with time and location encountered by Earth orbiting and balloon-borne experiments. 

\section*{Acknowledgment}

The authors acknowledge: Laurent Desorgher for providing an updated version of the PLANETOCOSMICS software compatible with Geant 4.9.5; the Swedish National Space Board for funding; Alex Howard and Christoffer Lundman for valuable discussions on this project; Miranda Jackson and Elena Moretti for comments on the manuscript; anonymous referees for providing detailed comments on this paper. The data from the Oulu Neutron Monitor were provided by Sodankyl\"a Geophysical Observatory via \url{http://cosmicrays.oulu.fi/}.





\end{document}